# Frequency dependent polarisation switching in h-ErMnO$_3$

*Alexander Ruff[1], Ziyu Li[1], Alois Loidl[1], Jakob Schaab[2], Manfred Fiebig[2], Andres Cano[2], Zewu Yan[3,4], Edith Bourret[4], Julia Glaum[5], Dennis Meier[5], Stephan Krohns[1,\*]*

1 Experimental Physics V, Center for Electronic Correlation and Magnetism, University of Augsburg, Augsburg 86159, Germany.
2 Department of Materials, ETH Zurich, 8093 Zürich, Switzerland.
3 Department of Physics, ETH Zurich, 8093 Zürich, Switzerland.
4 Materials Sciences Division, Lawrence Berkeley National Laboratory, Berkeley, California 94720, USA.
5 Department of Materials Science and Engineering, Norwegian University of Science and Technology, 7030, Trondheim, Norway.

\* stephan.krohns@physik.uni-augsburg.de

**We report an electric-field poling study of the geometric-driven improper ferroelectric h-ErMnO$_3$. From a detailed dielectric analysis we deduce the temperature and frequency dependent range for which single-crystalline h-ErMnO$_3$ exhibits purely intrinsic dielectric behaviour, i.e., free from extrinsic so-called Maxwell-Wagner polarisations that arise, for example, from surface barrier layers. In this regime ferroelectric hysteresis loops as function of frequency, temperature and applied electric fields are measured revealing the theoretically predicted saturation polarisation in the order of 5 - 6 µC/cm$^2$. Special emphasis is put on frequency-dependent polarisation switching, which is explained in terms of domain-wall movement similar to proper ferroelectrics. Controlling the domain walls via electric fields brings us an important step closer to their utilization in domain-wall-based electronics.**

Manipulation of ferroelectric (FE) domains constitutes an interesting research field aiming at the application of domains and domain walls (DWs) as functional elements in future nanoelectronics.[1,2,3,4,5] Among other multiferroic materials, special emphasize is currently put on the hexagonal manganites $R$MnO$_3$ ($R$ = In, Y, Er, Dy, Ho … Lu), because of their unusual ferroelectric properties. The hexagonal manganites display geometrically driven improper ferroelectricity ($T_c$ ≈ 1200 – 1500 K)[6] with six possible domain states, forming vortex-like topological defects.[7,8] The density of vortices strongly depends on the cooling rate as demonstrated in refs. [9,10]. The corresponding improper ferroelectric DWs, which connect the topologically protected vortices,[7,8,11] are explicitly robust and exhibit different functional properties, such as anisotropic electronic conductance and the local emergence of electronic inversion layers.[2,3,5] As demonstrated by Han *et al.*[12] in h-ErMnO$_3$ using high-resolution TEM experiments under electric fields, polarisation switching is a feasible way to control the shape and position of domains and domain walls, respectively. Polarisation switching experiments for hexagonal manganite, however, are still rare and comprehensive frequency and temperature dependent studies are highly desirable in order to understand the fundamentals of their macroscopic response to electric fields. So far, insight was mainly gained from FE hysteresis loops recorded at fixed temperature and frequency, as well as pyrocurrent measurements.[7,12,13,14,15,16,17,18] In particular, it is important to note that barrier layer capacitances[13,14] arising, e.g., from the formation of Schottky diodes at the sample-contact interface can superimpose the intrinsic dielectric response at ambient temperatures.[3,7] Hence, measurements of the intrinsic non-linear dielectric response, which are often performed in a two-point contact mode at ambient temperature, are hampered due to the so-called extrinsic Maxwell-Wagner (MW) relaxation of barrier layers.[19] As known for many oxide materials[20,21,22] a thorough dielectric investigation accompanied with an equivalent circuit analysis[20] is crucial for determining the appropriate temperatures and frequencies at which the *intrinsic* bulk dielectric properties are measured. Recently, this methodology was applied for YMnO$_3$ evidencing barrier layer capacitances for various samples at ambient temperatures.[14]

In this work, we focus on the h-ErMnO$_3$ system, which shows spontaneous polarisation of about 5 – 6 µC/cm$^2$ below 300 K[16] as well as switchable FE domains.[12] In contrast to so-called proper ferroelectrics, such as BaTiO$_3$, the polarisation is a secondary order parameter being driven by a unit-cell tripling structural transition as discussed, e.g. in refs 16 and 23. The associated domain pattern containing topologically protected vortex-like defects hampers the poling into a single-domain state.[7,8] The latter is a signature of the improper nature of the FE order in h-ErMnO$_3$, visible in spatially resolved measurements. To what extent this impacts on the macroscopic switching behaviour needs to be verified. As comprehensive studies of the bulk ferroelectric properties of h-ErMnO$_3$ are rare, we investigate the polarisation reversal in the bulk sample employing dielectric spectroscopy and hysteresis loop measurements. For the poling experiments a thorough frequency (1 Hz to 2 MHz) and temperature (50 K to 300 K) dependent dielectric analysis for h-ErMnO$_3$ is provided. Subsequently, FE hysteresis loop measurements are performed for various frequencies (0.1 Hz to 1 kHz) and at lower temperatures ruling out barrier-layer contributions – these hamper poling experiments at ambient temperatures – to the overall dielectric response. We investigate the polar switching behaviour of this improper FE material as function of electric field, frequency of the applied excitation field and temperature. The predicted value of the FE polarisation of 5 - 6 μC/cm$^2$ is confirmed.[16,23] While most previous studies have been focused on the static properties, Yang *et al.*[24] recently calculated the dynamic FE hysteresis loops. In the present study we measure now that dynamic behaviour and compare it with the theoretical prediction. Our results support DW movement, similar to proper ferroelectrics, as the sole mechanism of polarization switching.

A high-quality single crystal of h-ErMnO$_3$ was prepared via the pressurized floating zone technique, as described in detail by Yan *et al.*[25] The crystal was oriented by Laue diffraction and cut into disc-shaped sample. The sample had lateral dimensions of about 2 – 3 mm and a thickness of about 80 μm, exhibiting out-of-plane polarisation (P||c). The as-grown domain pattern was investigated using a commercial atomic force microscope (Ntegra Prima from ND-MDT), revealing a vortex density of about 10$^4$/mm$^2$ (or 0.01/μm$^2$). For the electrical measurements silver paint was used as metal contact on the top and bottom side of the plate-like sample. We performed dielectric spectroscopy using a Novocontrol Alpha analyser (frequency range from 1 Hz to 2 MHz) and FE hysteresis loop measurements using a TF2000 Aixacct system in combination with a high voltage booster for external electric fields up to 2 kV. For cooling the sample from 50 K to 300 K the measurements were conducted in a closed-cycle refrigerator and in vacuum to avoid electrical breakdown.

Figure 1 shows the frequency dependent dielectric constant ε' (a) and conductivity σ' (b) of h-ErMnO$_3$ for various temperatures. ε'(ν) for 300 K reveals a distinct step-like decrease from "colossal"[26] values of up to 10$^4$ at low frequencies (< 1 kHz) to intrinsic values of about 15 at higher frequencies (> 10 kHz). For lower temperatures ε' shows a plateau at approximately 10$^3$, which shifts with decreasing temperatures to lower frequencies. For T < 150 K no step remains in ε' within the measured frequency range from 1 Hz to 1 MHz. In addition, σ'(ν) indicates a step-like increase from a low conductivity to a plateau, which denotes the bulk dc conductivity σ$_{dc}$. For example, for σ'(250 K) this plateau is established between 100 Hz and 10 kHz exhibiting σ$_{dc}$ in the order of 10$^{-8}$ Ω$^{-1}$cm$^{-1}$. For higher frequencies, the bulk conductivity shows a power-law increase indicating the regime of universal dielectric response.[27]

These step-like features in the dielectric properties are typical signatures of MW relaxations often arising in electrical heterogeneous samples from internal or surface barrier layer capacitances (SBLC).[22] Here, the formation of Schottky barriers most likely leads to depletion layers at the sample surfaces, constituting thin barrier layers. These layers act as thin capacitors superimposing the intrinsic dielectric properties of the bulk material. However, as discussed in a recent work,[14] besides the SBLC effect, internal barrier layers can be formed by insulating DWs giving rise to a second step-like feature. In general, to obtain quantitative information we use an equivalent circuit model[14] fitting the dielectric spectra (lines in figure 1a and b), which allows separating the dielectric properties of the bulk and the barrier layers. The revealed temperature dependent σ$_{dc}$ of the bulk is marked as crosses in the inset of figure 1 denoting an Arrhenius law with activation energy of 0.29 eV, which is in excellent agreement with values reported in literature.[14,28] In addition, and consistent with the framework of a barrier layer mechanism,[20] σ$_{dc}$ follows the left flank of the relaxation feature arising in σ'(T) for various frequencies.

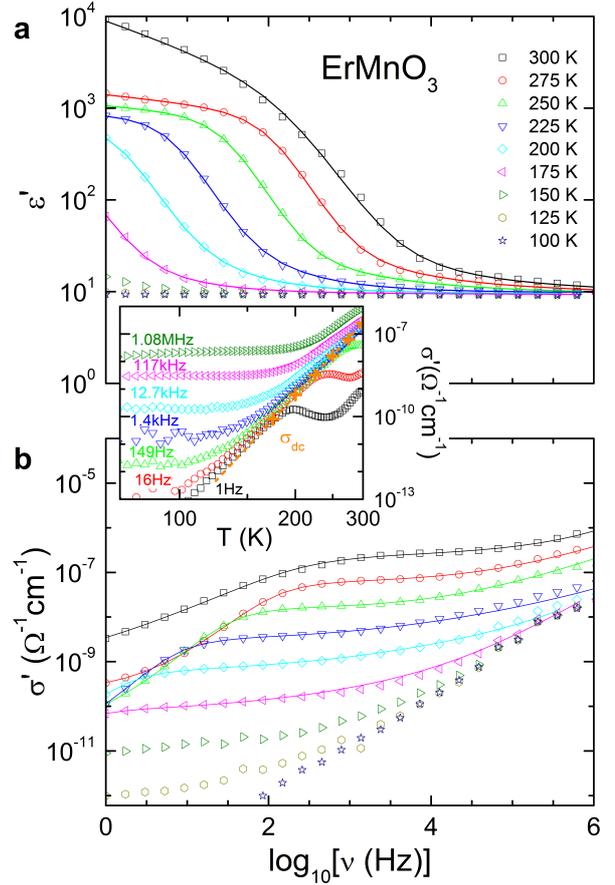

FIG 1. **Frequency dependent dielectric properties of h-ErMnO$_3$.** Dielectric constant (a) and conductivity (b) for various temperatures; sample oriented with out-of-plane polarisation. The lines are fits with an equivalent circuit, as reported by Ruff *et al.*[14] The inset denotes the Arrhenius representation of the conductivity featuring a thermally activated behaviour of the bulk conductivity.

The detailed knowledge about the (ν,T)-behaviour of the barrier layer in h-ErMnO$_3$ is crucial, because it can strongly affect non-linear dielectric measurements and even lead to artificial ferroelectric-like hysteresis loops.[19,29] Thus, in order to avoid parasitic, non-ferroelectric contributions, polarisation switching analysis was performed in the frequency range of 0.1 Hz to 1 kHz at temperatures below 160 K. Pure FE polarisation reversal, almost not affected by barrier-layer contributions, is for the present sample possible at low temperatures (T < 160 K) and sufficiently high electric activation fields $E_a$ exceeding the coercive field $E_c$, which is in the order of 40 kV/cm for the sister compound $YMnO_3$.[7] Figure 2 shows ferroelectric hysteresis loop measurements as function of temperature (a), applied excitation field (b) and frequency of the excitation field (c) for h-$ErMnO_3$. The polarisation pattern is imaged by piezo-response force microscopy (PFM) with scans obtained at different surface positions on both sides of our disc-shaped sample. In all scans we observe elongated ferroelectric domains as reflected by the representative PFM image in Fig. 2d. The domains exhibit a lateral extension of about ~ 0.5 – 3 µm in a predominantly stripe-like pattern, which differs from the almost isotropic domain pattern of h-$ErMnO_3$ within refs. [7,10]. The electric field dependent measurements of the dielectric constant (figure 2e) and the loss tangent (= ε"/ε') (figure 2f) reveal the intrinsic dielectric properties at 120 K (c.f. figure 1) as well as a peak-like feature that occurs while switching the polarisation. The latter is a common feature in proper ferroelectrics.[30] Increasing the temperature leads to an additional peak-like feature partially merging and superimposing the original peak of polarisation reversal. As already discussed for temperatures exceeding 120 K, barrier layers start to play a role up to the point that they dominate the non-linear dielectric response.

As a consequence, well-defined hysteresis loops are detected for 10 Hz and applied electric fields of 114 kV/cm (figure 1a), exhibiting a saturation polarisation $P_s$ in the order of 5 – 6 µC/cm$^2$ for temperatures around 120 K. For increasing temperature (T > 150 K) the hysteresis loop changes shape due to the aforementioned incipient barrier layer contributions. This is expressed in P(E) as additional dynamic processes during the polarisation reversal, resulting in two different slopes around $E_c$ as well as a curved saturation polarisation around the highest applied electric field. Such features are absent at 120 K and 10 Hz, where the barrier layer has no impact on P(E). For these parameters, maximum polarisation reversal requires that $E_a$ exceeds at least 2.5 |$E_c$|, which is illustrated in figure 2b. A lower applied $E_a$, or an increased $E_c$ for lower temperatures (e.g., T < 120 K shown in figure 2a), results in partial switching and hence a reduced value measured for $P_s$. This regime may also denote the creep regime of FE domain-wall motions.[31] For the chosen frequency range from 0.1 Hz to 1 kHz, 120 K is a suitable temperature to study the detailed frequency-dependent polarisation-reversal process (figure 1c). The P(E) loops show systematic increases of $E_c$ for higher applied frequencies, which again results in a reduced value of the measured $P_s$. These frequency-dependent P(E) loops are in perfect agreement with the theoretically predicted ones – calculated using the temporal evolution of order parameters derived from a time-dependent Ginzburg-Landau equation – of Yang et al.[32] The calculated P(E) loops are solely derived from movement of DWs connecting topologically protected vortices. Within this framework of pure DW motion, the time-dependent kinetics of the DWs determine the frequency-dependence of the polarisation reversal at higher frequencies. Here, we provide for h-$ErMnO_3$ temperature dependent quantitative data of the frequency dependent polarisation reversal and coercive fields evidencing this theoretical model.

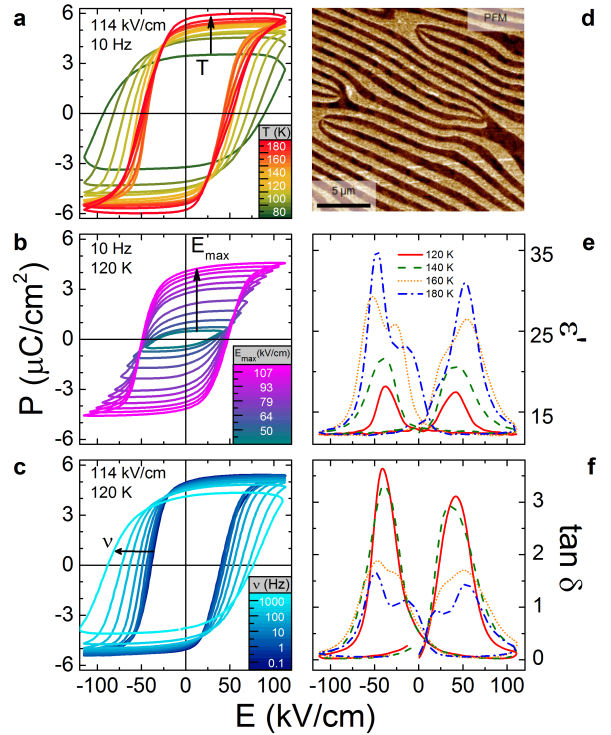

FIG 2. Electric field poling of improper ferroelectric order. Temperature- (a) electric-field- (b), and frequency- (c) dependent hysteresis loop measurement of a polarised h-$ErMnO_3$ single crystal. (d) polar surface pattern via piezo-response force microscopy. Electric field dependent dielectric constant (e) and loss tangent (f) for various temperatures.

In Figure 3 we show an analysis of the frequency dependent logarithm of the coercive field (a) for various temperatures as well as frequency and temperature dependent diagrams of $E_c$ (b) and $P_s$ (c). Only data are shown that exhibit mainly intrinsic dielectric behaviour. Ishibashi and Orihara[33] derived from the Kolmogorov-Avrami-Ishibashi model, which is often used describing the switching kinetics in proper ferroelectrics, a more simplified scenario for systems with deterministic nucleation of polar domains. In this model, the DW motion leads to domain growth, thus being solely responsible for polarisation reversal.[33] Within that framework, the volume fraction of reversed polarization depends on the frequency of the applied field and its waveform (usually sinusoidal). In the case of P(E) for h-ErMnO$_3$ the preferred domains expand and the opposite domains shrink into narrow stripe-like domains, as they cannot simply disappear. The latter is related to the topologically protected vortex-antivortex pattern, which is formed as discussed in detail in refs [7,8]. Theoretically, as described in ref. [24], the two boundary DWs of the stripe-like domains can merge into an unstable broadened wall, which splits again into the two afore merged DWs when reducing the external applied field. The observed remnant polarisation denotes a balanced state of preferred domains as well as opposite stripe-like domains. This remnant polarisation is stable in time at low temperatures but strongly decreases if the system is heated up to ambient temperatures. If $E_a$ exceeds the quasi-static threshold (c.f. figure 2c, $E_c$ remains constant for $\nu(E_a) < 3$ Hz at 120 K), the kinetics of DW motions contribute. In the scope of the Ishibashi-Orihara model, $E_c$ follows a simple power law: $E_c \propto \nu^\beta$. Linear fits yield an empirical β-exponent in the order of β = 0.103±0.0133 for temperatures ranging from 70 K to 160 K. This is quite similar to β-exponents of pure DW motion in conventional ferroelectrics, like lead zirconate titanate (β = 0.05)[34] and strontium bismuth tantalate (β = 0.12)[35] as well as multiferroic DW motion of LiCuVO$_4$ (β = 0.08)[36]. This corroborates the pure DW motion as mechanism for polarisation reversal as proposed by Yang et al.[24] Our results from the thorough ferroelectric hysteresis loop analysis are summarized in detailed diagrams of the coercive field showing a thermally activated behaviour as well as the frequency dependent kinetics of DW motion. As a consequence of the thermally activated behaviour, $E_c$ linearizes in the double logarithmic scale. The hatched area denotes the quasi-static regime. For this frequency and temperature dependent parameter set and in the case of $E_a$ = 114 kV/cm, the maximum of the saturation polarisation is achieved, which is shown as dark blue area in figure 3c. So, the saturation polarisation and consequently the corresponding DW motion are strongly influenced by temperature as well as the frequency and the value of the excitation field.

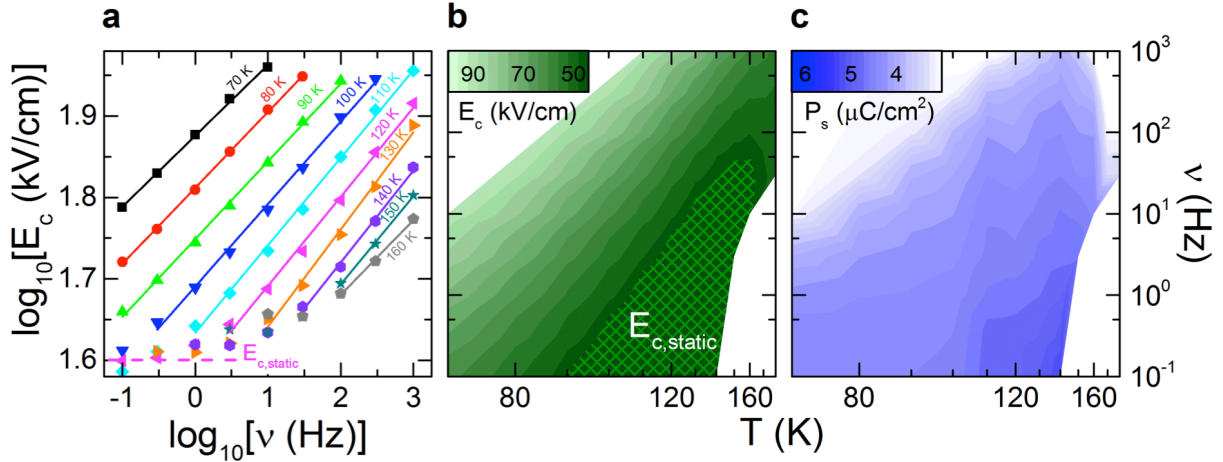

**FIG 3. Domain wall contributions to ferroelectric polarisation.** (a) Frequency dependence of the coercive field in double-logarithmic scale for various temperatures and driving frequencies. Lines are linear fits following the Ishibashi-Orihara theory. Temperature and frequency dependent phase diagrams of (b) coercive field and (c) saturation polarisation.

In conclusion, semiconducting materials like $R$MnO$_3$ often show barrier layer mechanisms superimposing the dielectric response of the intrinsic ferroelectric polarisation. A detailed systematic dielectric analysis is a prerequisite for a thorough ferroelectric characterisation. We clearly reveal a regime, where intrinsic properties become accessible. This allows investigating the polarisation reversal of single-crystalline h-ErMnO$_3$, which is a prime example for improper ferroelectric hexagonal manganite. Using a simplified model based on Ishibashi-Orihara theory for DW movement in proper ferroelectrics, we determine the frequency-dependent evolution of the coercive field, which follows a simple power law. The achieved power-law exponent is in perfect agreement to values of proper ferroelectrics showing pure domain wall movement. This approach enables the control of DWs in the improper ferroelectric manganites and brings us a step closer towards their utilization in domain-wall based electronics.[2,5]

**Acknowledgements**
A.R., Z.L., A.L. and S.K acknowledge funding from the DFG via the Transregional Collaborative Research Center TRR 80 (Augsburg/Munich/ Stuttgart, Germany), the "Fürstlich und Gräflich Fuggersche Stiftung" and from the BMBF via ENREKON 03EK3015. D.M. thanks NTNU's Onsager fellowship program for support. M.F. thanks for support by the ETH Research Grant and the ERC Advanced Grant INSEETO (No. 694955). M.F., D.M., J.S. acknowledge funding from ETH Zurich and SNF (proposal no. 200021_149192). J.G. acknowledges the EU call H2020-MSCA-IF-2014 under grant number 655866. Crystals were grown at the Lawrence Berkeley Laboratory supported by the U. S. Department of Energy, Office of Science, Basic Energy Sciences, Materials Sciences Division (Contract No. DE-AC02-05-CH11231).